\documentclass{WileyMSP-template}

\begin{document}

\pagestyle{fancy}
\rhead{\includegraphics[width=2.5cm]{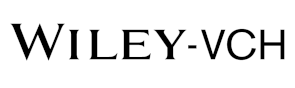}}

\title{Generalized Mie theory for full-wave numerical calculations of scattering near-field optical microscopy with arbitrary geometries}

\maketitle


\author{D\'aniel Datz*}
\author{Gergely N\'emeth}
\author{L\'aszl\'o R\'atkai}
\author{\'Aron Pekker}
\author{Katalin Kamar\'as}


\dedication{}

\begin{affiliations}
D\'aniel Datz, Dr. Gergely N\'emeth, Dr. \'Aron Pekker, Dr. L\'aszl\'o R\'atkai, Prof. Katalin Kamar\'as\\
Institute for Solid State Physics and Optics, Wigner Research Centre for Physics,
Konkoly-Thege Miklós út 29-33, 1121, P.O. Box 49, H-1525 Budapest, Hungary
Email Address: datz.daniel@wigner.hu

Prof. Katalin Kamar\'as
Institute of Technical Physics and Materials Science, Centre for Energy Research 
P.O. Box 49, H-1525 Budapest, Hungary

Dr. Gergely N\'emeth
SOLEIL Synchrotron, L'Orme des Merisiers, RD128, Saint Aubin, France 91190

\end{affiliations}



\begin{abstract}

Scattering-type scanning near-field optical microscopy is becoming a premier method for the nanoscale optical investigation of materials well beyond the diffraction limit. A number of popular numerical methods exist to predict the near-field contrast for axisymmetric configurations of scatterers on a surface in the quasi-electrostatic approximation. Here, a fully electrodynamic approach is given for the calculation of near-field contrast of several scatterers in arbitrary configuration, based on the generalized Mie scattering method. Examples for the potential of this new approach are given by showing the coupling of hyperbolic phonon polaritons in hexagonal boron nitride layers and showing enhanced scattering in core-shell systems. In general, this method enables the numerical calculation of the near-field contrast in a variety of strongly resonant scatterers and is able to accurately recreate spatial near-field maps.

\end{abstract}


\section{Introduction}

Scattering-type scanning near-field optical microscopy (s-SNOM) has become one of the leading methods for determining local optical information of materials with spatial resolution well below the diffraction limit. This method is especially effective in visualizing and otherwise investigating exotic optical phenomena, such as plasmon and phonon polaritons in 2D van der Waals crystals \cite{dai2019phonon, babicheva2018near} and detecting strong coupling between hexagonal boron nitride (hBN) and nanotube plasmons \cite{nemeth2022direct} or molecular vibrations \cite{li2018boron, bylinkin2021real, dolado2022remote}.

The extreme spatial focusing and amplification of the illuminating light happens by approaching the (often metallized) probing tip of an atomic force microscope (AFM) to close proximity of the investigated sample.
In this electromagnetic environment the AFM tip acts as an antenna and scatters light in every direction. The complex scattering processes between the sample/substrate and the tip slightly modify the scattering character of the probe. This small change in the amplitude and the phase of the scattered light can be detected by interferometric techniques, such as the heterodyne or the pseudo-heterodyne method \cite{ocelic2007phd}.

The exact nuances of the tip-sample interaction are still not entirely understood. A number of methods exist, with different complexity levels, that try to capture the essential details of the scattering process. Full-wave calculations with large program packages, such as finite element modeling (FEM) with COMSOL \cite{mester2020subsurface, chen2021rapid, chen2022rapid} or finite difference time domain (FDTD) \cite{kotlyar2021dual} include all the complexities in exchange for largely increased computational time and effort.
Simpler models, such as the point dipole model (PDM) \cite{raschke2003apertureless}, the finite dipole model (FDM) \cite{cvitkovic2007analytical} or the the extended finite dipole model (EFDM) \cite{cvitkovic2009phd} provide a quasi-static approximation to the solution of the scattering problem. In the PDM, the AFM tip is approximated by a point dipole, or a sphere with much smaller effective radius than the exciting laser wavelength. In the FDM and EFDM spheroidal scatterers are included for more realistic tip shape approximation. 
Lately, full-wave finite element calculations were combined with PDM and FDM formulations to achieve efficient extraction of the near-field contrast, albeit still in the quasi-static limit \cite{chen2021rapid, chen2022rapid}.

More involved quasi-static calculations include the polarizability of spheroidal and more complicated tip shapes and multilayer structures \cite{hauer2012quasi}. These models all severely underestimate the penetration depth which is important in the examination of multilayered thin films. 
A common drawback of these approximations is the use of fitting parameters with questionable physical interpretation to adjust the calculated results to measured ones. 

More sophisticated numerical methods without \textit{ad hoc} fitting parameters have also been developed for the calculation of the scattered signal in the fully electrodynamical limit.
The so-called "lightning rod" model \cite{mcleod2014model} does give a full description of electrodynamic effects, such as retardation effects, by using a method similar to the one presented in this paper.

Reference \cite{jiang2016generalized} uses the generalized spectral method for the description of the near-field contrast in case of highly resonant samples and realistic tip shapes.

While these models are able to calculate the near-field contrast for realistic tip shapes in the fully electrodynamic limit, they require translational symmetry in the plane (not accounting for the tip), which limits their applications regarding multiple scatterers in arbitrary configuration.

In this paper, we present a full-wave, tractable, relatively time-efficient method of calculating complex far-field scattering amplitudes and phases related to s-SNOM measurements, without spurious fitting parameters, using the "generalized Mie scattering" or multipole reflection theory (MRT) method.

\section{Methods}

Mie's theory yields the scattering properties of a sphere suspended in a homogeneous, non-absorbing medium by expanding the electromagnetic field in terms of spherical vector functions. The generalization of the classical Mie theory involves the inclusion of multiple, possibly layered and non-spherical scatterers in the vicinity of a plane interface representing the surface of a possibly layered substrate. 

Following the MRT method \cite{borghese1984multiple, fucile1997general, fucile1997optical, denti1999optical, borghese2013superposition, borghese2007scattering}, the fields can be expanded in terms of the spherical multipoles $\textbf{J}^{(p)}_{lm}$ and $\textbf{H}^{(p)}_{lm}$. The extension of the scattered electric field takes the form 

\begin{equation} \label{eq:esexp}
    \mathbf{E}_{sca} = 
    \sum_{\eta} E_{sca,\eta}
    \sum_{plm} \mathbf{H}^{(p)}_{lm}(\mathbf{r}')
    A^{(p)}_{\eta lm},
\end{equation}

where $\eta$ is the index for the different polarizations in a given polarization basis, and $A^{(p)}_{\eta lm}$ are the expansion coefficients. The index \textit{p} is either 1 or 2 and distinguishes between electric and magnetic type of vector functions. The indices \textit{l} and \textit{m} are the azimuthal and magnetic indices, respectively.

In the presence of multiple scatterers, the usual boundary conditions on the surface of each scatterer (the continuity of the electric and the magnetic field) result in a system of linear equations for the expansion coefficients:

\begin{equation} \label{eq:linear}
    \sum_{p'l'm'} M^{(pp')}_{lm,l'm'} A^{(p')}_{\eta l'm'} = -\mathcal{W}^{(p)}_{\eta lm},
\end{equation}

where the matrix \textit{M} is the matrix that describes the boundary conditions, while the coefficients $\mathcal{W}^{(p)}_{\eta lm}$ are the expansion coefficients of the exciting field.

In the presence of a plane interface, the reflected spherical multipoles are described by the formulation of Bobbert and Vlieger \cite{bobbert1986light}. Solving Equation \ref{eq:linear} for the expansion coefficients of the scattered field allows the formulation of the scattering amplitude matrix that connects the incident electric field to the observed electric field scattered in a given direction

\begin{equation} \label{eq:scamat}
    \mathbf{E}_{obs}(\theta,\phi) = \mathbf{f}(\theta,\phi) \cdot \mathbf{E}_{inc}.
\end{equation}

This equation is analogous to the commonly cited relation between the near field and the far field in SNOM literature ($E_N = \sigma E_I$) and makes the scattering amplitude matrix the final derived quantity of this formulation. 

The scattering parameters of objects of varying shapes can be given by the extended boundary condition method (EBCM). This method does not limit the shape of the particle as long as the surface is parametrizable \cite{mishchenko1999light, mishchenko2009electromagnetic}.

Details about the calculations can be found in the supporting information.

\section{Results}

\subsection{Phonon polariton coupling in hexagonal boron nitride}

Hexagonal boron nitride  is a naturally occurring hyperbolic material \cite{dai2014tunable}. This type of anisotropy causes the bulk phonon polaritons in the crystal to propagate on the surface of a cone. For the near-field detection of these polaritons, the phonon polaritons have to be coupled into the crystal by sharp features of the surface, such as the SNOM tip itself \cite{dai2014tunable}, the crystal edge or other objects on the surface \cite{dai2017efficiency} that are able to provide the missing momentum components for the coupling. The s-SNOM method is then able to visualize the phonon polaritons by mapping the fringe pattern arising from the interference of the polaritons coupled in by the tip of the SNOM device and the sharp feature. 

The main strength of the MRT formulation is the ability to handle additional particles besides the probing tip. To illustrate the power of this method, we calculated the phonon polariton interference fringes in a thin layer of hexagonal boron nitride (hBN). A gold nanosphere is placed on the surface of the hBN layer to couple phonon polaritons into the layer that can interfere with the tip-launched polaritons.

\begin{figure}
  \includegraphics[width=\linewidth]{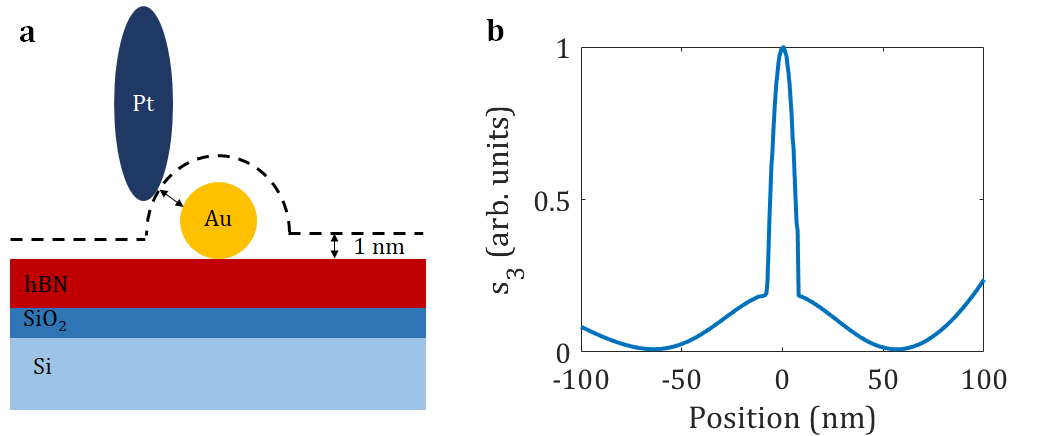}
  \caption{(a) Modeled geometry of a gold sphere and platinum tip above the multilayered Si/SiO$_2$/hBN surface. (b) Calculated near-field amplitude (third harmonic) acquired along the dashed line in (a).}
  \label{fig:goldsph}
\end{figure}

The layered structure consists of a 40 nm thick hBN layer on top of a 4 nm SiO$_2$ layer on a semi-infinite silicon substrate (see Figure \ref{fig:goldsph}a). The refractive index of the 3 nm radius gold sphere is extrapolated from measured data of Johnson and Cristi \cite{johnson1972optical}. The SNOM tip is modeled as a platinum spheroid of 600 nm length and 20 nm tip-inscribed sphere radius. The refractive index of platinum is also extrapolated from measured data \cite{rakic1998optical}. The illumination is a p-polarized plane wave with 60$^\circ$ angle of incidence. The calculations were conducted at 1540 cm$^{-1}$ excitation. A measurement value (amplitude and phase) is obtained by imitating the vibration of the tip by calculating at different separations between the tip and the sample. All the steps of the pseudo-heterodyne detection technique are replicated in the background-free extraction of the near-field amplitude and phase.

The near-field amplitude, demodulated at the third harmonic, is shown in Figure \ref{fig:goldsph}b. It clearly shows a high contrast peak at the position of the gold nanosphere caused by the high refractive index of gold. On either side of the sphere peak, the amplitude shows a characteristic ripple pattern. The approximate wavelength of this pattern (120-130 nm) agrees reasonably well with measured data \cite{menabde2022near}. This shows the ability of the generalized Mie scattering model to capture the propagation of phonon polaritons in multilayered surfaces containing hBN.

\begin{figure}
  \includegraphics[width=\linewidth]{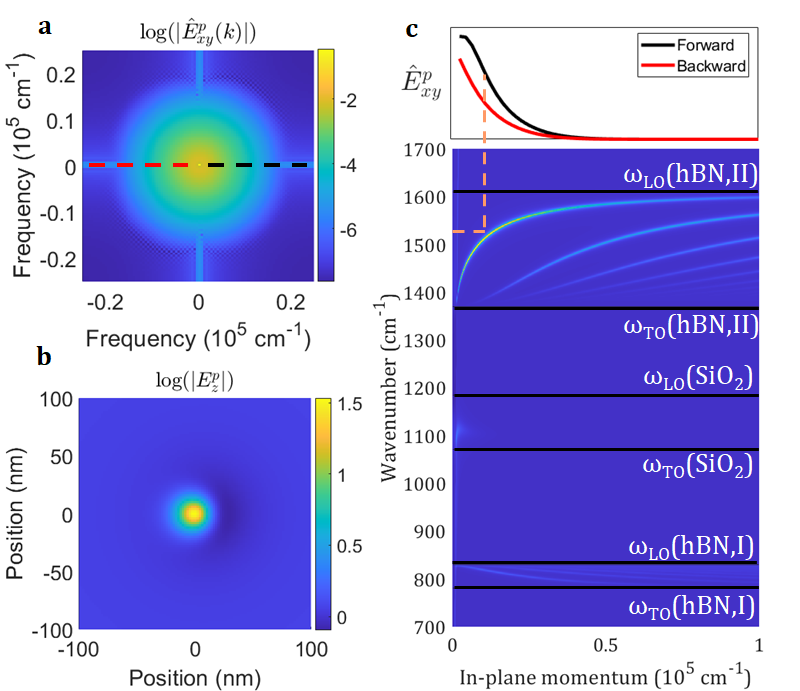}
  \caption{(a) Magnitude of the in-plane Fourier components of the near field above the surface of the multilayered structure in the presence of the platinum tip. Black and red dashed lines indicate the forward and backward scattering directions respectively. (b) Amplitude of the z component of the electric field (logarithmic scale), showing the nanofocus and the asymmetry of the field distribution. (c) (Below) Imaginary part of the reflection coefficient of the multilayered structure. (Above) The Fourier components in the forward (black) and backward (red) scattering directions.}
  \label{fig:hbnmomentum}
\end{figure}

A feature of the calculated linescan in Figure \ref{fig:goldsph} is the enhanced forward scattering of phonon polaritons marked by the larger amplitude in the forward scattering direction. This can be understood by analyzing the electric field on the surface of the multilayer structure. In Figure \ref{fig:hbnmomentum}a, the in-plane components of the electric field are visualised on the surface of the substrate in the presence of only the SNOM tip. 

The asymmetry of the in-plane field distribution in the forward and backward scattering direction (see Figure \ref{fig:hbnmomentum}b for the z component) results in significantly different amplitude Fourier components for the propagating modes in the multilayer pattern. In Figure \ref{fig:hbnmomentum}c the available Fourier amplitudes are depicted for 1540 cm$^{-1}$ excitation, showing the larger in-plane amplitude in the forward scattering direction.

\subsection{Enhanced scattering from encapsulated molecules in boron nitride nanotubes}

\begin{figure}
  \includegraphics[width=\linewidth]{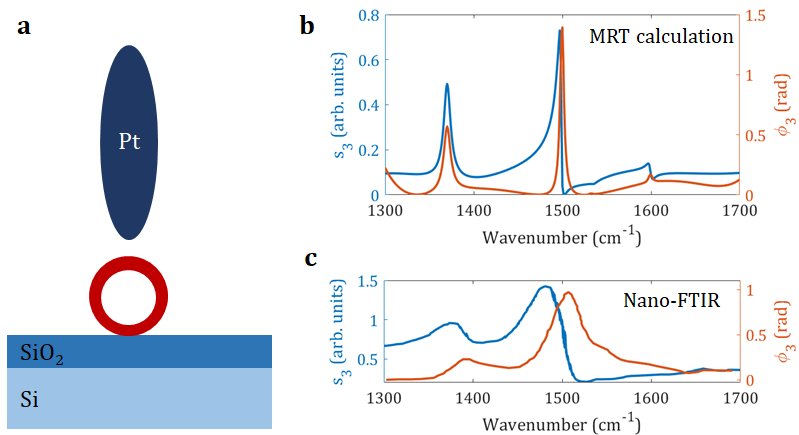}
  \caption{MRT calculations of the layered sphere representing a BNNT (a) Schematic geometry of the system.  (b) Calculated spectrum with peaks at 1380 cm$^{-1}$ and 1500 cm$^{-1}$. The 1380 cm$^{-1}$ peak comes from a separate calculation of an hBN layer of the same thickness as the BNNT wall.  (c) Extracted nano-FTIR spectrum of 20 nm thick BNNT from \cite{phillips2022mid}. The peak positions match well with the calculated results. The observed broadening of the measured peaks can be caused by the significantly larger diameter of the BNNTs.}
  \label{fig:bnnt_hollow}
\end{figure}

The near-field optical properties of boron nitride nanotubes (BNNTs) are exciting due to the nanotubes' ability to enhance the scattering of the encapsulated material inside its cavity \cite{datz2021polaritonic}. Using the generalized Mie scattering model, the approximate numerical modeling of BNNT scattering is possible. For simplicity, the BNNT is modeled as a hollow sphere (3 nm outer, 2 nm inner radius) with its refractive index set to that of hBN's out-of-plane refractive index (see Figure \ref{fig:bnnt_hollow}a). The substrate is 4 nm SiO$_2$ on top of a semi-infinite silicon half-space. The calculation parameters otherwise coincide with the ones mentioned in the previous section.

The calculated near-field amplitude and phase (third harmonic) are shown in Figure \ref{fig:bnnt_hollow}b. Since the scattering properties of anisotropic spheres are hard to calculate numerically \cite{roth1973scattering}, the particle under the SNOM tip is practically a hollow, metallic shell. Since the average diameter of BNNTs is rather large (5-20 nm diameter), the TO phonon polariton peak at 1370 cm$^{-1}$ coincides well with the same peak in hBN. The results in Figure \ref{fig:bnnt_hollow}b are thus the sum of the results from the hollow sphere and from an hBN layer of the same thickness as the BNNT wall thickness. These results are compared to nano-FTIR measurements reported in Reference \cite{phillips2022mid}. The calculated spectra show reasonable agreement for the position and the symmetry of the peaks. The difference of the peak widths can be attributed to the large difference in diameter between the measured nanotubes and the one used for the calculation.

An important conclusion drawn from the calculated results is that the BNNT peak at around 1500 cm$^{-1}$ is caused by the characteristic Mie scattering of hollow nanospheres and not by any additional polaritonic excitation inherent to only boron nitride materials. This example shows that the generalized Mie scattering method is able to reproduce the near-field spectral features of more complicated geometries and give meaningful information about the origin of the peaks.

\begin{figure}
  \includegraphics[width=\linewidth]{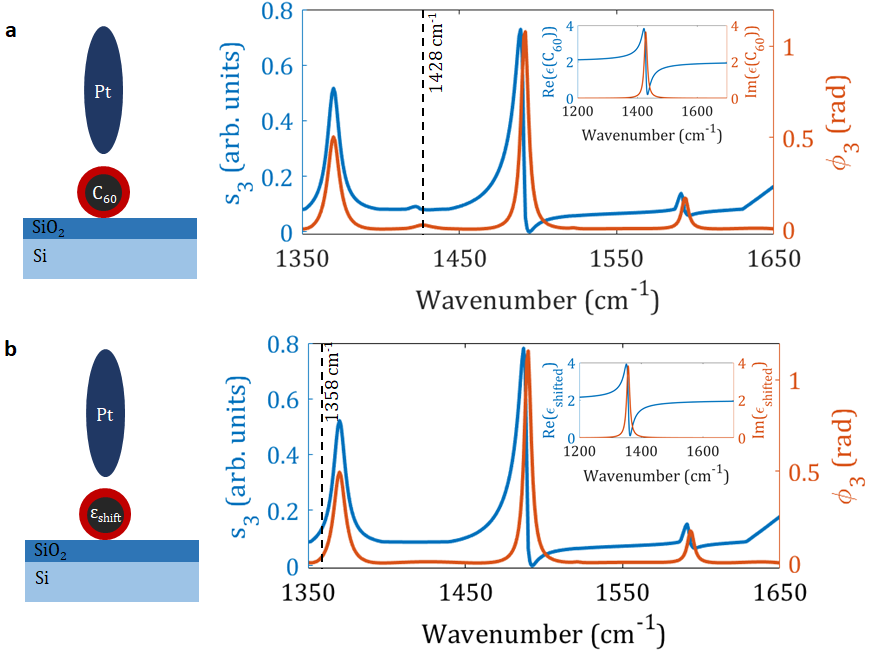}
  \caption{MRT calculations of layered spheres representing filled BNNTs. (a) The resonance frequency of the dielectric function of the inner sphere is set to 1428 cm$^{-1}$ (see inset). A small peak is visible at this wavenumber. (b) The resonance frequency of the dielectric function is set to 1358 cm$^{-1}$. The additional peak from the inner sphere disappears.}
  \label{fig:bnnt_filled}
\end{figure}

In Reference \cite{datz2021polaritonic}, we showed that the signal of weakly absorbing C$_{60}$ molecules can be detected inside BNNTs due to the enhancing effect of the highly confined electromagnetic fields inside the walls of the nanotube. This configuration can also be handled with the MRT model, using core-shell layered spheres representing the nanotubes. The inner sphere's refractive index is given by a single Lorentzian with parameters describing a weak oscillator. In Figure \ref{fig:bnnt_filled}a, the resonance frequency of the inner sphere is set to 1428 cm$^{-1}$ which coincides with a $T_{1u}$ vibrational mode of C$_{60}$ \cite{pusztai1999bulk, kovats2005structure, klupp2007vibrational}. Using the MRT method, a small peak in both the near-field amplitude and phase is present, the shape of which is typical for weak vibrations. On the other hand, if the resonance frequency of the Lorentzian of the inner sphere is shifted to 1358 cm$^{-1}$ (see Figure \ref{fig:bnnt_filled})b) which approximately coincides with a $(C_{60})_3$ trimer vibration mode \cite{stepanian2006ir} and is outside of the Reststrahlen band, this effect disappears. The results show that the field enhancement in the inner sphere predicted by Mie's theory of a core-shell spherical system \cite{xie2015efficient} is enough for a significant enhancement of the near-field signal. In a more realistic configuration, the additional field confinement due to the hyperbolic nature of the nanotube can result in even higher electric field amplitude which can explain the amplitude of the detected peaks in Reference \cite{datz2021polaritonic}.

\section{Conclusion}

In this paper, we introduced a novel approach to calculate near-field contrast in s-SNOM measurements using a generalized Mie theory approach, the multipole reflection theory. This model is able to calculate the near-field contrasts of several, complex scatterers in arbitrary configuration in a fully electrodynamic treatment. The scattering parameters of all of the interacting particles are calculated using Mie's theory, therefore the use of arbitrary fitting parameters is unnecessary.

Using the MRT model, we showed that it is suitable to calculate hyperbolic polariton interference fringes coupled into a thin hBN layer by a nanoparticle on its surface. Furthermore, the polaritonic enhancement of scattering from the inner sphere of core-shell nanoparticles can also be calculated that provides valuable insight into the enhancement of molecular vibrations inside BNNTs. 

Further improvements to the presented numerical model are still possible. The scattering properties of disks (\cite{kristensson1982t}) and cylinders (\cite{yan2008scattering}) are possible to calculate numerically and can be included into the model without a great increase in complexity. The illumination is also better described by a Gaussian beam \cite{bareil2013modeling} rather than a plane wave. The results presented in this paper are the first steps in accurate numerical modeling of non-symmetric geometries with multiple particles using the scattering theory approach.

\medskip
\textbf{Supporting Information} \par 
Supporting Information is available from the Wiley Online Library or from the author.

\medskip
\textbf{Acknowledgements} \par 
We gratefully acknowledge the support from the National Research, Development and Innovation Office - NKFIH FK-138411 and K-143153. Research infrastructure was provided by the Hungarian Academy of Sciences (MTA).

\medskip

%
\bibliographystyle{MSP}
\bibliography{bibliography}

\end{document}